\documentclass{aa}
\usepackage{txfonts}
\usepackage{graphicx}

\newcommand{\oiii}{[\ion{O}{iii}]}

\begin{document}
\title{A peculiar planetary nebula candidate in a globular cluster in 
       the Fornax dwarf spheroidal galaxy
  \thanks{Based on observations collected at the European Southern
          Observatory, Chile under programme 078.B-0631(A), and
	  with the NASA/ESA Hubble Space Telescope.
	  }
}

\author{S{\o}ren S. Larsen}

\institute{
  Astronomical Institute, University of Utrecht, Princetonplein 5,
  NL-3584 CC, Utrecht, The Netherlands
  \email{larsen@astro.uu.nl}
}

\offprints{S.\ S.\ Larsen, \email{larsen@astro.uu.nl}}

\date{\today}

\abstract
{
  Until now, only one planetary nebula (PN) has been known in the Fornax dwarf spheroidal galaxy.
}
{
  The discovery of a second PN candidate, associated with 
one of the 5 globular clusters in the Fornax dwarf, is reported.
}
{
  Spectra of the globular cluster H5, obtained with the UVES
echelle spectrograph on the ESO Very Large Telescope, show \oiii\ line
emission at a radial velocity consistent with membership of the Fornax
dwarf.  A possible counterpart of the \oiii\ emission is
identified in archival images from the Wide Field Planetary Camera 2 on
board the Hubble Space Telescope.  The source of the emission is located about 
$1\farcs5$ (less than one core radius) southwest of the centre of the cluster. 
}
{
The emission line source is identified as a likely PN, albeit with
several peculiar properties. No H$\beta$, \ion{He}{i}, or \ion{He}{ii} line 
emission is detected and the \oiii /H$\beta$ ratio is $>25$ (2$\sigma$).
The expansion velocity inferred
from the \oiii\ 5007\AA\ line is $V_e \approx 55$ km s$^{-1}$, which is large
for a PN.  The diameter measured on the HST images is about $0\farcs23$
or 0.15 pc at the distance of the Fornax dSph. 
}
{
This object doubles the number of known PNe in Fornax, and is only the 5th
PN associated with an old GC for which direct imaging is available. It may
be a member of the rare class of extremely H-deficient PNe, the second
such case found in a GC.
}


\titlerunning{A peculiar planetary nebula in the Fornax dSph}
\maketitle

\section{Introduction}

Like many other dwarf spheroidal galaxies, the Fornax dwarf spheroidal
(dSph) has a remarkably complex star formation history. Colour-magnitude
diagrams show evidence of star formation until as recently as $\sim10^8$ 
years ago (Stetson et al.\ \cite{stet98}; Buonanno et al. \cite{buon99}). 
The metallicity distribution of the field stars peaks at [Fe/H]$\sim-1.0$,
but with tails extending up to nearly solar metallicity and down to 
[Fe/H]$\approx-2.7$ (Helmi et al.\ \cite{hel06}).  
Battaglia et al.\ (\cite{bat06}) identified three more or less distinct 
stellar populations in the Fornax dSph, namely an old, metal-poor component, 
a dominant intermediate-age (2--8 Gyr) moderately metal-rich component,
and a younger ($\sim 1$ Gyr) metal-rich component.

Five globular clusters (GCs) are known in the Fornax dSph, catalogued
by Hodge (\cite{hod61}) and hereafter referred to as H1\ldots H5. When
normalised 
to the (low) luminosity of the galaxy, this leads to an extremely high 
GC specific frequency of $S_N\approx 29$ (e.g.\ van den Bergh \cite{van98}).
Historically, it is interesting to note that
one of the GCs (NGC 1049 = H3) was already discovered by John Herschel 
nearly a century before the parent galaxy itself.  From high-dispersion 
spectroscopy of individual stars in clusters H1, H2, and H3, 
Letarte et al.\ (\cite{let06}) 
derived [Fe/H]=$-2.5\pm0.1$, [Fe/H]=$-2.1\pm0.1$, and [Fe/H]=$-2.4\pm0.1$, 
respectively.  From low-resolution spectroscopy, Strader et al.\ 
(\cite{stra03}) found an average [Fe/H]$\sim-1.8$ and old ages
(comparable to Milky Way GCs), with H5 possibly a few Gyrs younger than
the rest.  Thus, the GCs are 
most likely associated with the old, metal-poor component in Fornax.

The Fornax dSph is known to host a single planetary nebula (PN), discovered 
by Danziger et al.\ (\cite{dan78}).  In a recent study of this object, 
Kniazev et al.\ (\cite{knia07}) derived an oxygen abundance of 
12+log(O/H) = $8.28\pm0.02$ ([O/H] = $-0.55$ for the 
solar oxygen abundance of Grevesse \& Sauval \cite{gs98}). They also 
estimated an iron abundance of [Fe/H] = $-1.13\pm0.18$. It seems likely that 
this PN is associated with the intermediate-age component, but as for 
most PNe, pinpointing the exact age (and hence the mass of the
progenitor star) is difficult. 

This \emph{letter} reports the discovery of a second PN candidate in 
the Fornax dSph, apparently associated with one of the GCs (H5). This 
not only doubles 
the number of known PNe in Fornax, but also constitutes a significant addition 
to the small number of PNe currently known to be associated with GCs in general 
(four in Milky Way GCs; Jacoby et al.\ \cite{jac97}). The identification of 
PNe in stellar clusters holds special significance since it gives a 
better handle on the age and metallicity of the progenitor star (although
binary star evolution may be an important path for PN formation in clusters).
Furthermore, the Fornax dSph is close enough that spatially 
resolved imaging of the PN is within reach.

\section{Identification of a new PN candidate in H5}

\subsection{UVES spectroscopy}

  Three of the Fornax GCs (H3, H4, and H5) were observed with
the UVES echelle spectrograph on the ESO Very Large Telescope on Nov
19 and Nov 20, 2006. The primary aim of the observations
was detailed abundance analysis from the integrated cluster 
light, which will be described elsewhere.  To sample the \emph{integrated} 
light of the clusters, the UVES slit was scanned across each cluster in the 
east-west and north-south direction. In each direction, two scans of 2400 s 
each were obtained, which for H5 covered
a range of $\pm10\arcsec$ with respect
to the cluster centre (the half-light radius of H5 is $9\farcs6$; 
Mackey \& Gilmore \cite{mg03}).
The observations were carried out with the
red arm alone, covering a wavelength interval of 4200\AA -- 6200\AA\ with
a small gap near 5200\AA\ between the two CCD detectors.  The slit length
and width were $8\farcs9$ and $1\farcs0$, respectively, yielding 
a spectral resolution of $\lambda/\Delta\lambda\approx40,000$. The
observations were made under clear conditions with only occasional
thin clouds.

\begin{figure}
\includegraphics[width=85mm]{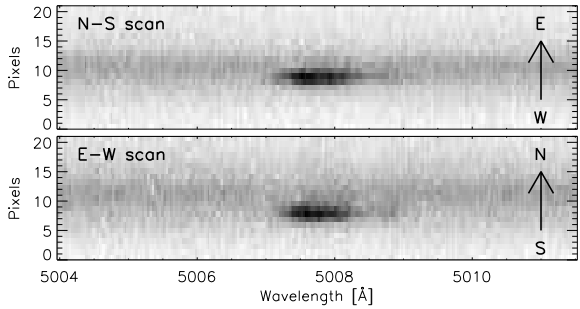}
\caption{Sections of the echellograms near the \oiii\ 5007\AA\ line.}
\label{fig:twodfig}
\end{figure}

Figure~\ref{fig:twodfig} shows the co-added east-west and north-south
scans of H5 for the spectral region near $5007$\AA . Each panel 
covers the full 
slit length of $8\farcs9$ with the arrows indicating the spatial
direction.  In addition to the stellar continuum from the cluster stars,
the scans clearly show emission at $\sim5008$ \AA, consistent with
the wavelength of \oiii\ at the radial velocity of the Fornax
dSph (53 km/s, according to the NASA/IPAC Extragalactic
Database). The emission is spatially unresolved in the UVES data
and offset from the centre of the GC trace by about 3.5 pixels 
($\sim1\farcs4$) towards the south and by about 2 pixels ($\sim0\farcs8$) 
towards the west. The source of the emission is thus located well within
one core radius ($2\farcs1$; Mackey \& Gilmore \cite{mg03}) from the
centre of H5.

\begin{figure*}
\includegraphics[height=170mm,width=63mm,angle=90]{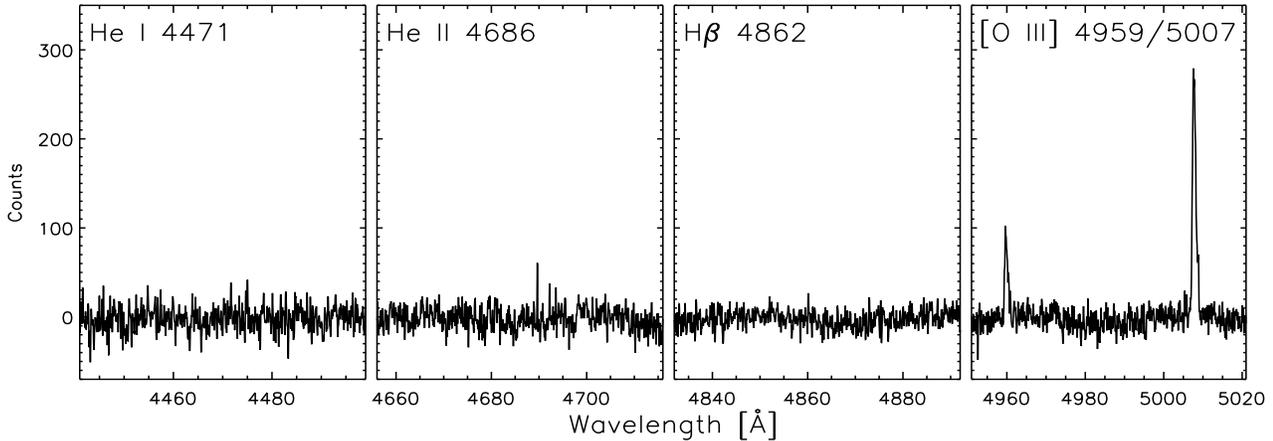}
\caption{Spectral regions around \ion{He}{i}, \ion{He}{ii}, 
 H$\beta$, and \oiii . The contribution from GC stars has been subtracted. 
 The spectra have been smoothed with a Gaussian of FWHM=5 pixels.  
Only the \oiii\ 4959,5007\AA\ lines are detected.}
\label{fig:specsub}
\end{figure*}

Figure~\ref{fig:specsub} shows one-dimensional spectra of the rows
with the line emission for selected wavelength intervals. These 
spectra were constructed by co-adding the rows containing line emission
and subtracting the sum of the remaining rows, scaled to 
the same mean level. An additional emission line, about 3 times fainter,
is seen at 4960\AA, confirming the identification of the line emission as 
\oiii . Although the two oxygen lines are detected at $\sim17\sigma$
and $\sim50\sigma$, no other emission features are seen in the spectrum.
In particular, there is no trace of either the H$\beta$ line, the 
\ion{He}{i} 4471\AA, or the \ion{He}{ii} 4686\AA\ lines.

Since no flux standards were observed, an accurate flux calibration of
the data is not possible and only crude constraints can be placed on
line fluxes and ratios. The observed count ratio of the two \oiii\
lines is $3.4\pm0.2$, close to the theoretically expected intensity ratio
of 3.0 (e.g.\ Osterbrock \cite{osterbrock}). Only upper limits can be
placed on the \oiii /H$\beta$ ratio: Since the \oiii\ 5007\AA\ line is
detected at about $50\sigma$, a 2$\sigma$ lower limit is
\oiii /H$\beta$ $>$ 25. This ratio only refers to uncalibrated
counts, but since the separation in wavelength between H$\beta$ and
\oiii\ is modest, the corresponding limit on the flux ratio is probably not 
very different. Even this lower limit is a very high \oiii /H$\beta$ ratio 
for a PN. The Acker et al.\ (\cite{ack92}) catalogue lists
\oiii /H$\beta$ line ratios for 886 Galactic PNe, with a median
\oiii /H$\beta$ ratio of 8.9 and only about 3\% of the PNe 
have \oiii /H$\beta$ $>$ 20. High \oiii /H$\beta$ ratios
usually imply a very high degree of excitation, but this should also be
accompanied by strong \ion{He}{ii} lines (Dopita \& Meatheringham \cite{dm90}),
which however remain undetectable here. An alternative explanation for the
lack of Balmer emission is depletion of hydrogen.

  A few K giants that were observed during the run can be
used to establish a crude flux calibration.  By combining 
our observations of HD~83516 and HD~223094 with their $B$ magnitudes,
a conversion between count rate and flux was defined
at $\sim4400$ \AA .  Applying this conversion also at 5000\AA, the 
corresponding \oiii\ 5007\AA\ line flux of the source in H5 is estimated to 
be $F_{5007} \approx 5\times10^{-15}$ ergs cm$^{-2}$ s$^{-1}$ or
$m_{5007} \approx 22$. The adopted conversion between count rate and
flux corresponds to a total throughput of $\approx10$\% (UVES + 
telescope + atmosphere) for an 8.2 m main mirror, in good agreement with 
the UVES on-line documentation.  The absolute magnitude is then 
$M_{5007} \approx 1$ for a distance of 137 kpc (Mackey \& Gilmore \cite{mg03}). 
In doing this calculation, it was taken into account that the slit 
acts as a ``shutter'' that scans $20\arcsec$ during each 2400 s
exposure, so that the effective integration time for any given
point on the sky is 120 s per scan.  The above 
numbers should be considered order-of-magnitude estimates only, but 
it does seem clear that this PN is relatively faint, about 6 mag
below the tip of the PN luminosity function (Jacoby \cite{jac89}).

\begin{figure}
\includegraphics[width=85mm]{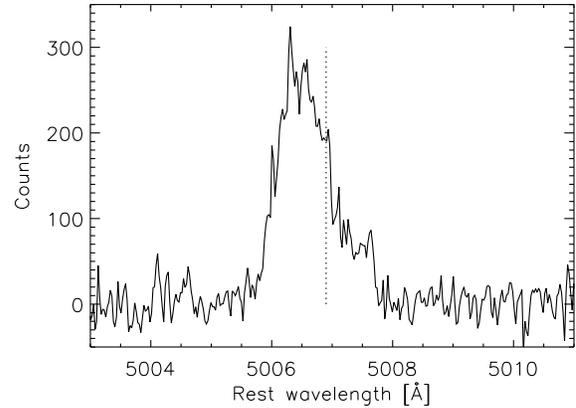}
\caption{The \oiii\ 5007\AA\ line profile. The contribution from GC
stars has been subtracted out and the wavelength scale corrected for
the radial velocity of H5. No smoothing has been applied.}
\label{fig:oiii}
\end{figure}

The \oiii\ line profiles are well-resolved in the
UVES spectra. Figure~\ref{fig:oiii} shows the profile of the
\oiii\ $\lambda$5007\AA\ line, corrected to zero radial velocity
using the H$\beta$ stellar absorption line seen in the GC spectrum.
The line profile extends from about 5005.9\AA\ to 5007.8\AA\ and is
strongly asymmetric, but the midpoint of this interval coincides
with the rest wavelength of \oiii\ (indicated by a vertical dashed
line). This further supports the physical association of the source of 
the \oiii\ emission with the Fornax dSph and H5.  The line half-width 
corresponds to an 
expansion velocity of $V_e \approx 55$ km/s, which is fairly high for a PN.
The median $V_e$ listed by
Acker et al.\ (\cite{ack92}) is 16 km s$^{-1}$, and only 4 of the 284 PNe 
in their compilation with $V_e$ measurements have
$V_e > 50$ km s$^{-1}$. 
Data for PNe in other Local Group galaxies confirm
that such high expansion velocities are rare and offer some evidence for a 
correlation of $V_e$ with excitation class,
although not with $M_{5007}$ (Richer \cite{rich06}).

\subsection{Possible identification of the nebula in HST/WFPC2 imaging}

To test whether the source of the \oiii\ emission seen in the UVES
spectra can be identified independently, images from the
Wide Field Planetary Camera 2 (WFPC2) on board the Hubble Space Telescope
were downloaded from the archive at the Canadian Astronomy Data Centre
and inspected. H5 was observed under programme 5917 (P.I.\ R.\ Zinn) 
in the F555W ($\approx$V) and F814W ($\approx$I) filters with exposure 
times of 5640 s and 7720 s.  The cluster is centred on 
the PC chip and the co-added images provided by the archive
were used directly.  

A section of the F555W image is shown in the left-hand panel of 
Fig.~\ref{fig:hstfig}, centred on the location of the emission line
source derived from the UVES spectra. The circle has a radius of
$1\arcsec$ and the arrows indicate the orientation (N up, E to the left).
The centre of the cluster is near the upper left corner of the
panel and identification of the PN counterpart is clearly complicated by
severe crowding. Several objects are located near the
expected location, but no unique identification of the \oiii\ source 
is immediately possible.

The ratio of the F555W and F814W images is shown in the right-hand panel of
Fig.~\ref{fig:hstfig}. Darker shading indicates
a higher ratio of F555W/F814W, i.e.\ a bluer integrated colour. Due to the
more extended point spread function in F814W, stars with red 
colours tend to appear as white annular structures, while stars with blue
colours appear as dark points.
Near the centre of the circle (slightly to the southeast) is an extended dark 
ring, which is identified as the likely counterpart of the \oiii\
emission.  The \oiii\ lines fall well within the transmission of the 
F555W filter, while no strong emission lines fall within
the F814W filter, so the nebula is indeed expected to have a blue 
F555W/F814W colour. Figure~\ref{fig:hstfig} hints at a slight 
asymmetry in the E-W direction, but generally the nebula appears fairly
round. The diameter is about 5 WFPC2 pixels, 
i.e.\ $0\farcs23$ or $\approx0.15$ pc, a fairly typical value
for PNe (e.g.\ Sabbadin et al.\ \cite{sab84}). 
The corresponding expansion age is then about 1300 years.

\section{Discussion and conclusions}

\begin{figure}
\begin{minipage}{42mm}
\includegraphics[width=42mm]{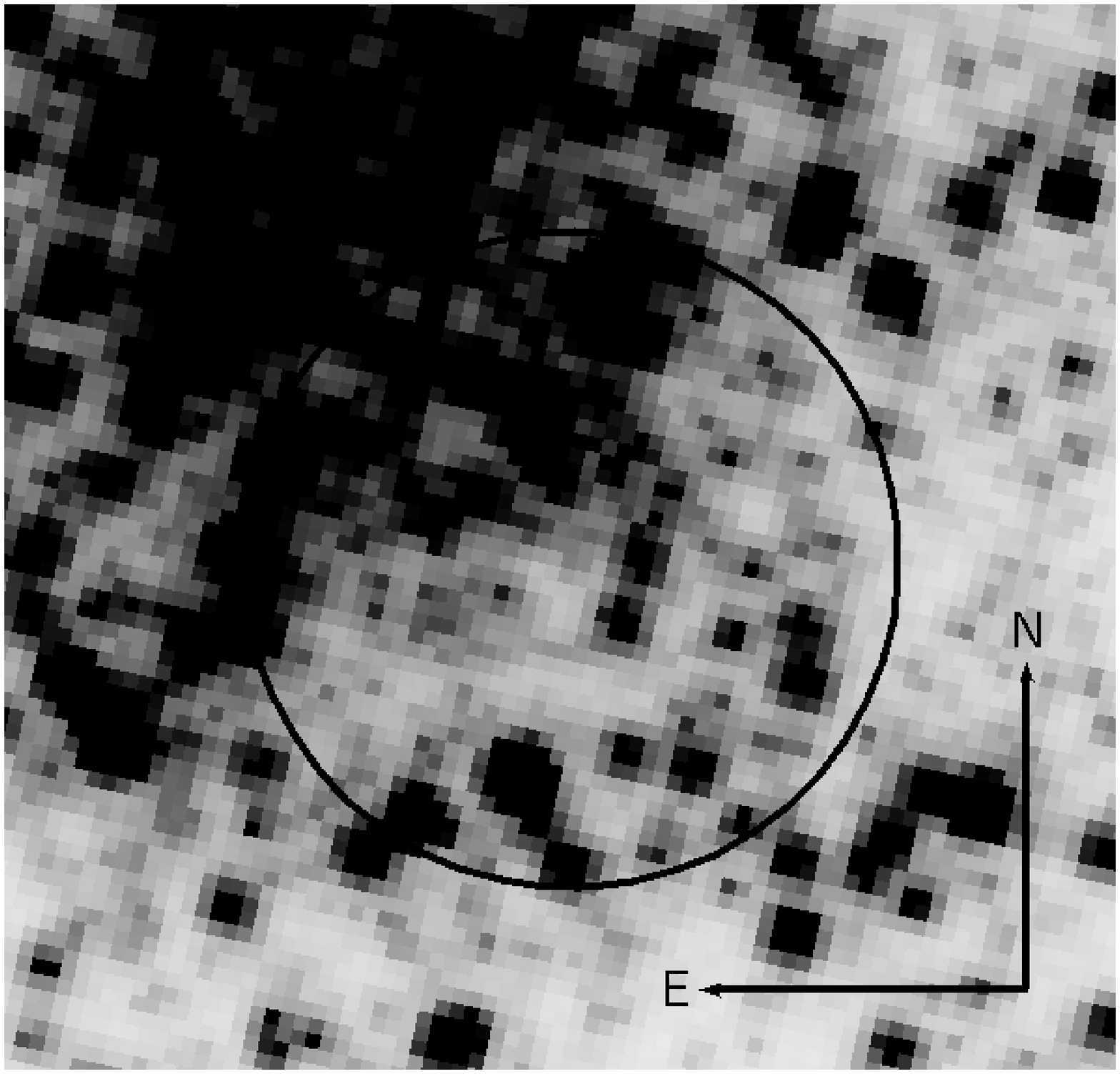}
\end{minipage}
\hspace{1mm}
\begin{minipage}{42mm}
\includegraphics[width=42mm]{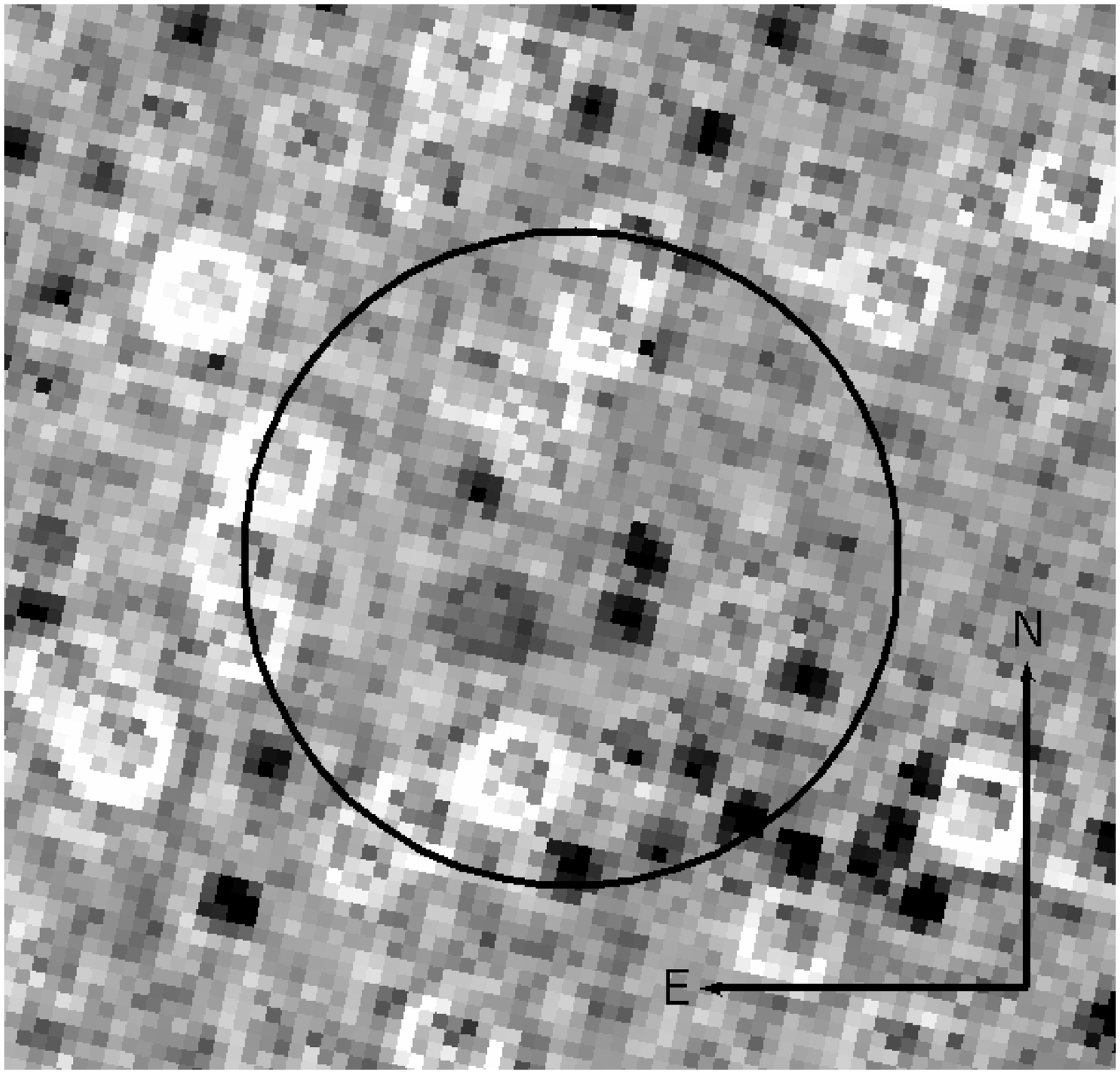}
\end{minipage}
\caption{Left: HST/WFPC2 F555W image of H5. Right: Ratio of
F555W and F814W exposures (black=higher ratio of F555W/F814W). The 
circle (diameter = 1$\arcsec$) is centred on the location of 
the emission line object derived from the spectra (Fig.~\ref{fig:twodfig}).
}
\label{fig:hstfig}
\end{figure}

The new PN candidate in H5 doubles the number of 
PNe in the Fornax dSph. That it has escaped detection so far, despite 
many studies of the Fornax GCs, is probably due to its relative 
faintness and location far enough from the centre ($\sim1\farcs5$) 
of H5 to be easily missed in an ordinary long-slit spectrum. As seen in
Fig.~\ref{fig:hstfig}, it is also inconspicuous even in deep (broad-band) HST 
images, although it should be easily identifiable in narrow-band WFPC2, 
ACS/HRC, or WFC3 images. 

As already noted by Danziger et al.\ (\cite{dan78}), comparison with other 
Local Group galaxies suggests that about one PN should be expected in the
Fornax dSph. More recently, Magrini et al.\ (\cite{mag03}) concluded from a 
census of PNe in 
the Local Group that one PN is expected per $10^{6.92} L_V/L_{V,\odot}$.
Assuming $M_V\approx-13.1$ for
Fornax (Demers et al.\ \cite{dem94}), 1.7 PNe are then expected.
Thus, the discovery of a second candidate does not 
change the conclusion that the PN population of Fornax is consistent
with what is observed in other Local Group galaxies. 
The Sagittarius dSph, which is somewhat brighter than Fornax,
is now known to host four PNe 
that span a range in metallicity and thus, like those in Fornax, appear 
to trace different populations (Zijlstra et al.\ \cite{zij06}). However,
none of these is associated with any of the $\sim8$ known GCs in the 
Sagittarius dSph (van den Bergh \cite{van07}).
What is the likelihood of detecting a PN in one of the Fornax GCs 
specifically?  Clearly, one should be careful about drawing statistical 
conclusions based on a sample of one object.  A naive scaling of the four 
PNe known in the $\sim150$ Milky Way GCs (Harris \cite{har96}), 
predicts about 0.13 PNe in the 5 Fornax GCs, but Poisson statistics
would still yield 1 PN in a Fornax-like GC system in about 11\% of the cases.

The nebula in H5 is a rather peculiar object. 
The expansion velocity is high for a PN, although hydrodynamical 
simulations (Villaver et al.\ \cite{vil02}) reproduce the observed 
combination of radius and $V_e$ at late stages of the 
PN evolution for a 1 $M_\odot$ progenitor. There is some other
evidence linking high PN
$V_e$ values 
to old stellar populations
(Richer \cite{rich06}). High $V_e$ values are also observed in 
bipolar PNe with intermediate-mass progenitors 
(Corradi \& Schwartz \cite{cs95}), but the nebula in H5 does 
not appear 
to belong to this class (Fig.~\ref{fig:hstfig}).
The non-detection of emission lines other than \oiii\ is even more
atypical.  In this regard the nebula resembles the peculiar 
PN IRAS 18333-2357 in 
the metal-poor Galactic GC M22, where the absence of emission lines other 
than \oiii\ 4959/5007\AA\ and [\ion{Ne}{iii}] 3869/3967\AA\ is thought to 
imply an extreme H depletion 
(Cohen \& Gillett \cite{cg89}).  
However, IRAS 18333-2357 is exceedingly faint: The \oiii\ flux is comparable 
to that of the nebula in H5, although M22 is a factor of 40 closer!
Extreme H depletion has also been found in knots within some Galactic
PNe, e.g.\ Abell 30 (Wesson et al.\ \cite{wes03}). A possible scenario is 
that this is the result of a late thermal pulse, occurring after 
most of the H-rich envelope had already been ejected (Iben et al.\ \cite{ib83}).
Depending on when exactly this happened, an outer low-surface brightness 
PN might also be present, as in Sakurai's object 
(Duerbeck \& Benetti \cite{db96}).

Further observations of the nebula in H5 
might help shed more light on its nature.  HST narrow-band imaging 
might be rewarding,
and a
comparison of the morphology with that of PNe in Galactic GCs and
in the Sagittarius dSph would be interesting, e.g.\ in relation
to differences in the motion with respect to an ambient 
interstellar/intergalactic medium.
The velocity dispersion of the
Fornax GC system is only 6 km s$^{-1}$ (Cohen \cite{coh83}),
but Fornax may still be interacting with intergalactic gas in the 
Local Group (although this point remains uncertain;
Piatek et al.\ \cite{pia07}).  
It would also be desirable to search 
for emission lines other than those from \oiii , such as [\ion{N}{ii}], 
[\ion{Ne}{iii}], ionised He, and (of course) the 
hydrogen Balmer lines.

\begin{acknowledgements}
J.\ Walsh, A.\ Zijlstra, O.\ Pols, H.\ Lamers, and the referee, Dr.
L.\ Magrini, are thanked for 
several useful suggestions. The abundance study for which the
UVES data were obtained is a collaboration with R.\ Peterson, J.\ Strader,
and J.\ Brodie of the SAGES group.  This research has used the facilities of 
the Canadian Astronomy Data Centre operated by the National Research Council 
of Canada with the support of the Canadian Space Agency.
\end{acknowledgements}

\end{document}